# SPATIAL TRENDS IN EDIACARAN BILATERIAN TRAILS

Brittany A. Laing[1,2], Luis A. Buatois[1], M. Gabriela Mángano[1], Glenn A. Brock[2,4], Zoe Vestrum[3], Luke C. Strotz[4], & Lyndon Koens[5]

## 5.1 ABSTRACT

The Savannah Hypothesis and the Cambrian Information Revolution invoke the development of spatially heterogeneous resource distribution during the Ediacaran-Cambrian transition as a key driver of infaunalization and sensory evolution of mobile bilaterians, respectively. However, difficulties in detecting historical resource distribution hinders the ability to tests these theories. If external conditions crucial to organism fitness (e.g. nutrient distribution, oxygen availability) became increasingly heterogeneous across the Ediacaran-Cambrian transition, then it follows that benthic organisms dependent on these conditions would demonstrate a similar increase in the spatial variability of their movement trajectories. To investigate Ediacaran resource distribution on the seafloor we examined the morphology of *Helminthoidichnites tenuis*, a simple unbranched horizontal bilaterian trail, from the Ediacara Member of Southern Australia for spatial trends. Our analysis reveals a as-yet undiscovered variability in the behaviour of the *Helminthoidichnites tenuis* tracemaker and confirmed heterogeneity in external conditions relevant to the tracemaker in the latest Ediacaran.

[1] Department of Geological Sciences, University of Saskatchewan, Saskatoon, Saskatchewan, Canada. [2]
[2] School of Natural Sciences, Macquarie University, Sydney, New South Wales, Australia.
[3] Department of Physics, University of Alberta, Edmonton, Alberta, Canada.
[4] State Key Laboratory of Continental Dynamics, Shaanxi Key Laboratory of Early Life & Environments and Department of Geology, Northwest University, Xi'an, China.
[5] Department of Mathematics, University of Hull, United Kingdom.

## 5.2 INTRODUCTION

The late Ediacaran period heralded the advent of animal motility. This can be observed in White Sea Assemblage trace fossils such as *Epibaion costatus, Kimberichnus teruzzi* and *Helminthoidichnites tenuis* (Fedonkin & Waggoner, 1997; Ivantsov & Malakhovskaya, 2002; Ivantsov, 2009; 2011; Gehling et al., 2014; Ivantsov et al., 2019; Bobrovskiy et al., 2018; Evans et al., 2018; 2020) and, to a lesser extent, in the shallow-marine deposits coeval with the Avalon Assemblage (Clarke et al., 2024). The succeeding Ediacaran-Cambrian transition records the onset of penetrative bioturbation documented by the appearance of ichnogenera such as *Treptichnus, Gyrolithes,* and *Psammichnites* (Mángano & Buatois, 2020). Subsequent infaunalization during the Fortunian-Cambrian Age 2 transition occurred concomitant with a first-order change in ecological structuring, from an Ediacaran-style matground ecology to a Cambrian-style mixground ecology (Mángano & Buatois, 2014; Gougeon et al., 2018), an evolutionary breakthrough known as the Agronomic Revolution (Seilacher, 1999).

Two hypotheses, the Savannah Hypothesis and the Cambrian Information Revolution, suggest the development of heterogeneous resource distribution was a key evolutionary driver. The Savannah Hypothesis suggests that increasingly spatially heterogeneous resource distribution, due to the presence of large biomass-rich organisms throughout the Ediacaran and Cambrian, facilitated infaunalization (Budd & Jensen, 2017). Similarly, the Cambrian Information Revolution posits that increased spatial heterogeneity, increased sensory capacity, and behavioural evolution formed a positive feedback loop throughout the Ediacaran and Cambrian with sophisticated information processing and a diversity of navigational devices already in place by the end of the Cambrian (Plotnick et al., 2010; Hsieh and Plotnick, 2022).

While direct measurement of resource distribution in the fossil record is difficult to obtain, the Ediacaran-Cambrian transition contains a prolific record of trace fossils. Specifically, there is an abundance of ichnogenera which predominately record combined movement and feeding (e.g. *Helminthoidichnites, Archaeonassa, Parapsammichnites, Psammichnites*). These fossil movement trajectories can be analyzed through the paradigm of movement ecology which describes the formation of a movement path as a function of the tracemaker's internal state, navigation capacity, and motion capacity as well as the external factors that affect movement (e.g. nutrient distribution, oxygen availability) (Nathan et al., 2008). As the expression of these factors change, so will the resulting optimal movement path. Simulations of optimal foraging paths can be conducted by dictating the expression of the four factors with explicit assumptions (e.g. the internal state is always "to find food", resources are patchily distributed). Likewise, movement paths can be recorded and used to hypothesize on the expression of the four factors (Codling et al., 2008). If external conditions crucial to organism fitness (e.g. nutrient distribution, oxygen availability) became increasingly heterogeneous across the Ediacaran-Cambrian transition, then it follows that benthic organisms dependent on these conditions should demonstrate a similar increase in the spatial variability of their movement trajectories.

Here, we develop a method to analyze spatial variability in movement trajectory morphology and apply this methodology on *Helminthoidichnites tenuis*, a simple unbranched horizontal trail, from the Ediacara Member of Southern Australia. *Helminthoidichnites tenuis* is interpreted as the trace of a bilaterian-grade organism, whose tracemaker in the Ediacara Member has been suggested to be *Ikaria warioota* (Evans et al., 2020). It is an unbranched, straight to curved, unlined, unornamented, and passively filled trail or burrow. Morphologic evidence for *H. tenuis* supports a grazing mode of life for the tracemaker, although scavenging on different macrofossils (e.g. *Dickinsonia, Aspidella, Funisia*) has been proposed by some authors (Droser et al., 2017; Gehling & Droser, 2018; Evans et al., 2020). Previous work on *Helminthoidichnites tenuis* from Ediacaran strata has indicated the presence of sensory-driven behaviours, making it a good candidate to examine spatial trends in external conditions via movement trajectories (Carbone & Narbonne, 2014; Paterson et al., 2017; Coutts et al., 2018; Gehling & Droser, 2018; Evans et al., 2020).

5.3 GEOLOGIC BACKGROUND

During the Proterozoic era and Cambrian period Australia was situated in low latitudes within the zone of carbonate development and formed the eastern margin of Gondwana (Pisarevsky et al. 2008; Brock et al., 2000). The strata deposited at this time in central and South Australia can be broadly categorized into two zones, the Centralian Superbasin (Officer, Amadeus, Ngalia, and Georgina basins, among others) and the Adelaide Fold Belt (Stansbury and Arrowie basins) (Munson et al., 2013). The 580-530 Ma Petermann Orogeny disrupted sedimentation in these basins; the uplift of the Musgrave province separated the Officer Basin from the Centralian Superbasin and the Adelaide Fold Belt was uplifted due to dextral shear (Zang et al. 2004). After a brief depositional hiatus extensive sedimentary deposition occurred over central and southern Australia in a vast quasi-contiguous depositional system spanning the Amadeus, Ngalia, southern Georgina, Warburton, Arrowie and Stansbury Basins (Munson et al., 2013). The formations of interest for this study are found within the Arrowie Basin, situated in the modern-day Flinders Ranges of South Australia. This basin formed on a rifted continental margin to the east of the Gawler Craton and records the deposition of a thick Neoproterozoic rift complex (Zang et al., 2004). Neoproterozoic sedimentation ended with the deposition of the Pound Subgroup, the youngest part of the Ediacaran Wilpena Group. The Pound Subgroup consists of the red siliciclastic rocks of the Bonney Sandstone which are disconformably overlain by the Chace Quartzite Member of the Rawnsley Quartzite. This member is unconformably overlain by the Ediacara Member, which is in turn conformably overlain by the upper Rawnsley Quartzite (Gehling, 2000; McMahon et al., 2020; Reid et al., 2020). The Rawnsley Quartzite encompasses a wide range of shallow- to marginal-marine environments, including tidally influenced rivers or estuaries, tidal flats, lagoons, shoreface, foreshore, backshore, and deltaic distributary channels, with body-fossil beds of the Ediacara Member interpreted to be fully marine and above fair-weather wave base (McMahon et al., 2020). The Ediacara Member is well-known for its exceptional preservation of iconic Ediacaran fossils, including *Aborea*, *Andiva, Dickinsonia, Ikaria, Kimberella, Parvancorina, Spriggina, Tribrachidium,* and *Yorgia* (Glaessner, 1958; Coutts et al., 2016; Evans et al., 2018; Xiao et al., 2020). Overlap of associated taxa with assemblages in the White Sea region of Russia provide a tentative age between 560 and 551 Ma (Evans et al., 2021 and sources therein). Among these body fossils occur the horizontal foraging trace fossil described as *Helminthoidichnites* isp. and herein referred to as *Helminthoidichnites tenuis* (Gehling & Droser 2018; Evans et al., 2021).

## 5.4 METHODS

We examined specimens of *Helminthoidichnites tenuis* located on four stratigraphic surfaces (i.e. slabs) from the Ediacara Member of Southern Australia (Figure 5.1). Photographs of each slab, taken as close to plan view as possible, were provided by Prof. Mary Droser. All discernable specimens were then traced in Adobe Illustrator with the pen tool, producing vector images of each specimen, with trail width and scale indicated. These images were subsequently imported into MATLAB and discretized according to the methodology outlined in Laing et al.

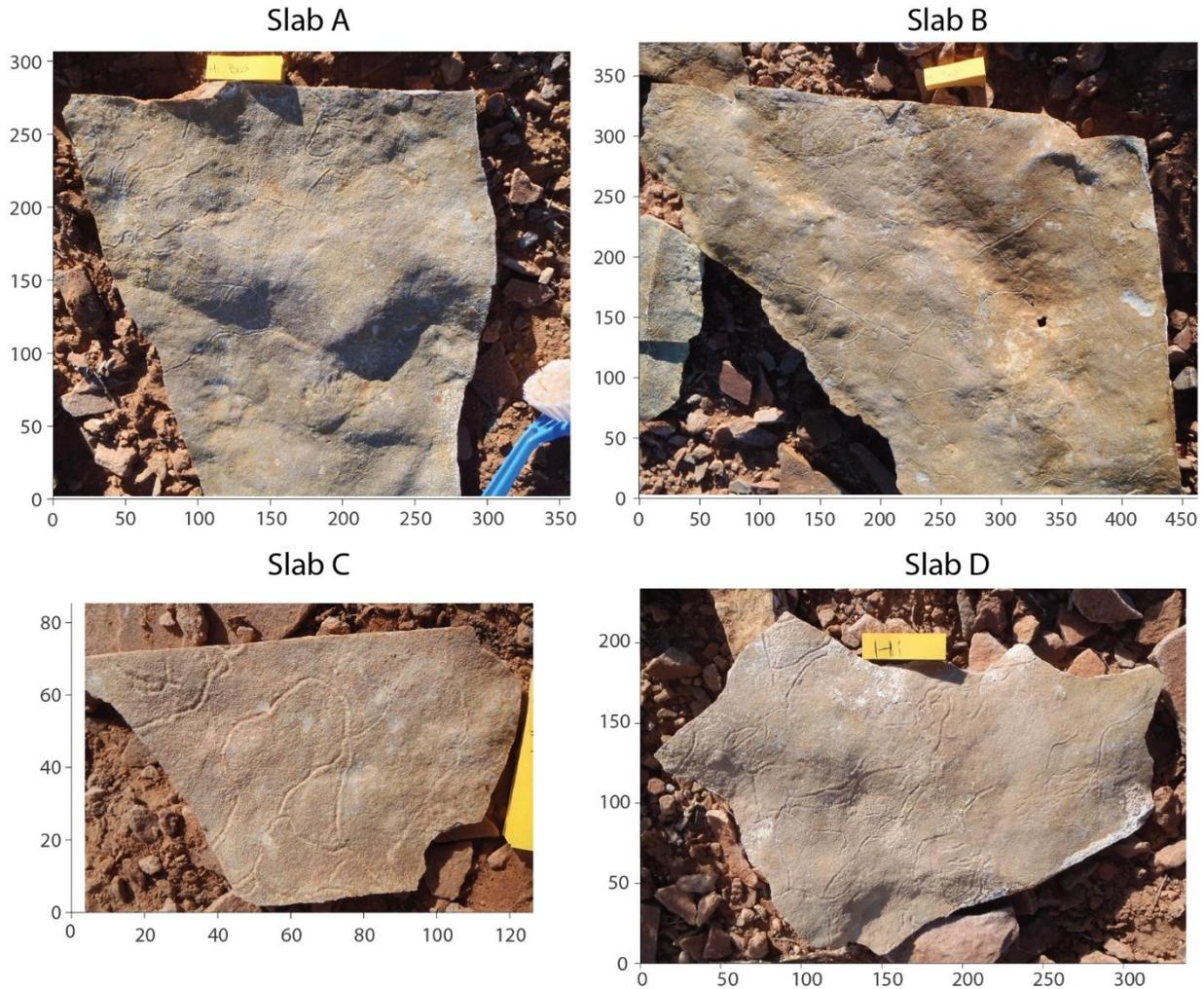

Figure 5.1. *Helminthoidichnites tenuis* bearing slabs from the Ediacara Member used in this study. Axes in mm.

(**Chapter 3**). This methodology converts images of trace fossil paths to 2D curves that can then be subdivided into equidistant segments along the fossil trajectory according to an inferred velocity distribution (Figure 5.2). For our specimens we assumed that an average velocity provided a reasonable approximation of the velocity distributions of each tracemaker. Each equidistant segment can be interpreted to represent the passage of an approximately constant unit of time, thus providing spatial and inferred temporal data of each trajectory. From this data descriptive measures can be calculated (Jones, 1977; Marsh & Jones,1988; Calenge et al., 2009; Dray et al., 2010).

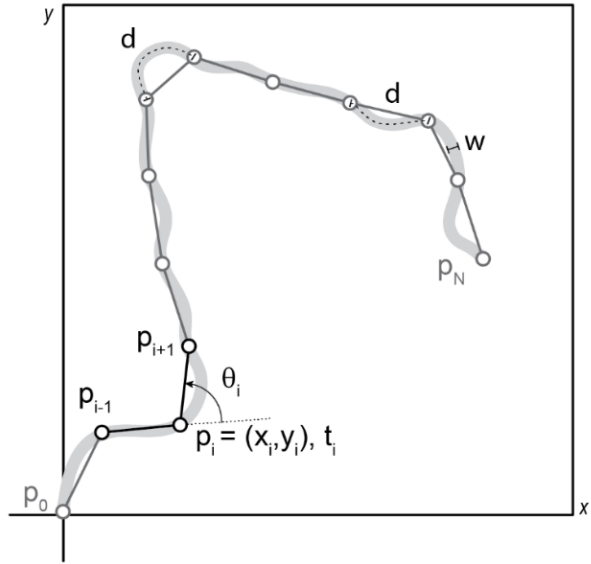

Figure 5.2. Sample movement path (thick grey line), subdivided into equidistant segments, with measures used in this study indicated: $p$ is the point data which delineates the discretized movement path (thin dark grey and black line), with associated $x$- and $y$- coordinates as well as inferred time ($t$) data, $\theta_i$ is the turning angle for the point $p_i$, $w$ is the trail width, $d$ (dotted black line) is the segment distance along the curvilinear length of the trail ($d = w*s$, where $s$ is the segment distance multiplier).

For our analyses, we calculated the relative turning angle taken at points ($p$) spaced equidistant along the trajectory. Each point represents the location of the organism after $i$ "steps". The distance between these points ($d$) is calculated by multiplying the trail width ($w$) by a segment distance multiplier ($s$). In this way the points are spaced according to a readily available biological measurement and is resilient to fluctuations in trail width within or across populations as well as possible errors in scale (**Chapter 3**). Prior analyses on the studied specimens demonstrated decreasing positive correlation between turning angles spaced smaller than 1.4 $w$ apart, which was interpreted to reflect the typical distance travelled in a single action by the organism (**Chapter 4**). As a result, we calculated the relative turning angles for each specimen at equidistant points ($p$) spaced 1.4 $w$ apart (Table 5.1).

Table 5.1. The number of specimens and phi values collected (at 1.4 $w$) for each slab.

| SLAB | NO. SPECIMENS | NO. PHI VALUES |
|---|---|---|
| **SLAB A** | 13 | 240 |
| **SLAB B** | 43 | 993 |
| **SLAB C** | 8 | 327 |
| **SLAB D** | 28 | 546 |
| TOTAL | **92** | **2106** |

Turning angles were then subsequently grouped and compared according to their spatial location on each slab. To do so, each slab was subdivided by an *x*-by-x grid with grid squares denoting spatial groupings of turning angles (Supp. Fig. 3). Analyses were run at a variety of grid spacings ($x = 5\ w$, $10\ w$, $20\ w$, $30\ w$, $40\ w$; where $w = 1.4$ mm) to determine the impact of grid size. It was determined that a grid spacing of $30\ w$ (42 mm) provided grid squares which often encompassed more than one specimen, at a resolution that would be reasonable for detectable environmental heterogeneity, including any heterogeneity imparted by the presence of *Dickinsonia costata* (mean length = 24.31 mm; Evans et al., 2017). Turning angles were subdivided into sample populations according to the grid square they belonged to, from which turning angle probability distributions (i.e. PDFs) were collected.

The probability distribution of turning angles offer a quantitative description of the morphology of a movement trajectory (**Chapter 3**). These distributions can be compared via statistical tests, such as two-sample *t*-tests or *f*-tests, to examine the likelihood that the population means (via a two sample *t*-test) or variances (via a two sample *f*-test) of two samples were equal (Supp. Fig. 2). Since our specimens show no consistent directional cues each turning angle was duplicated and multiplied by -1, symmetrizing the PDF's (i.e. left and right turns were not distinguished). This caused the means or each PDF to be 0 and precluded the use of two-sample *t*-tests. Instead, the variances of the spatially restricted turning angle sample populations from *Helminthoidichnites tenuis* trajectories were compared via two-sample *f*-tests to determine if the two samples differed significantly from each other. *P*-values, the results of these tests, describe the probability of obtaining a sample variance at least as large as the observed difference, assuming the samples came from populations with the same variances. Higher *p*-values, therefore, indicate a higher probability that the samples came from populations with the same variances. *P*-values are often discussed as a measure of the evidence against the null hypothesis ($H_0$), with smaller values indicating stronger evidence against $H_0$. One method to visualize the results of a series of two-sample *f*-tests is via a matrix of *p*-values (Supp. Fig. 2). In these matrices, each cell contains the *p*-value result obtained from a single two-sample *f*-test.

We developed a threshold argument to isolate grid squares whose overall movement patterns (i.e. turning angle distributions) were largely different to the majority of analyzed grid squares. Our argument included (1) a threshold *p*-value of 0.01, which we deemed to be sufficient to reject $H_0$ (i.e. significantly "different" enough) and (2) a threshold percentage of

139

33% of grid squares from which an individual specimen was significantly different from. For each grid square, we then summed how many *p*-values were greater than our threshold *p*-value and sorted our matrix of *p*-values by these summed amounts. Those grid squares not encompassed by our thresholding argument were deemed to represent regions where a dominant morphotype prevailed. Within the remaining grid squares, we observed a plaid or checkerboard patterning to the *p*-value matrix, with alternating regions with no evidence to disprove $H_0$ ($p > 0.1$) and regions with weak to strong evidence to disprove $H_0$ ($p < 0.1$). Working from the regions of no evidence to disprove $H_0$, we were able to further cluster the remaining grid squares into two groups, representing regions where two subordinate morphotypes prevailed (**Chapter 3**). Each grid square was then coloured according to their predominant morphotype (yellow for dominant, blue for low-variance subordinate, red for high-variance subordinate).

This analysis was performed twice; one for a grid localized at (0, 0) (i.e. the "unshifted" grid), and one for a grid localized at (15 *w*, 15 *w*) (i.e. the "shifted" grid). The results from both analyses were then superimposed for each slab, in effect dividing each slab by a 15 *w* by 15 *w* grid. Regions which revealed the same morphotype in both the "shifted" and "unshifted" analyses were outlined and indicated as candidate grid squares.

5.5 RESULTS

Our analyses reveals three movement patterns (i.e. morphotypes) within the Ediacara Member slabs which appear in spatially distinct regions (Figures 5.3 and 5.4). This includes one dominant movement pattern (yellow morphotype) and two subordinate movement patterns (blue and red morphotypes). This may reflect one dominant behaviour and two subordinate behaviours. Candidate grid squares for all three morphotypes were present on three out of the four slabs, of which 97 were of the dominant (yellow) morphotype, 44 of the low-variance subordinate morphotype (blue), and 30 of the high-variance subordinate morphotype (red). Of the bioturbated grid squares, 33% belonged to the dominant morphotype group, 13% to the low-variance subordinate morphotype group, 15% to the high-variance subordinate morphotype group, and 38% were unassigned (Table 5.2). Most candidate grid squares occurred directly adjacent to other grid squares of the same morphotype, forming large continuous candidate regions. While many candidate regions consist of only a few grid squares, five large candidate regions (> 5 grid squares, > 22.05 cm$^2$) of the subordinate morphotypes (3 low-variance, 2 high-variance), and six of the dominant morphotype were recognized. The largest candidate region encompasses an area of 48.51 cm$^2$. All candidate regions larger than three grid squares contain more than one specimen, indicating morphological consistency across specimens within these regions. In turn, many specimens display more than one morphotype, indicating morphological variation within specimens.

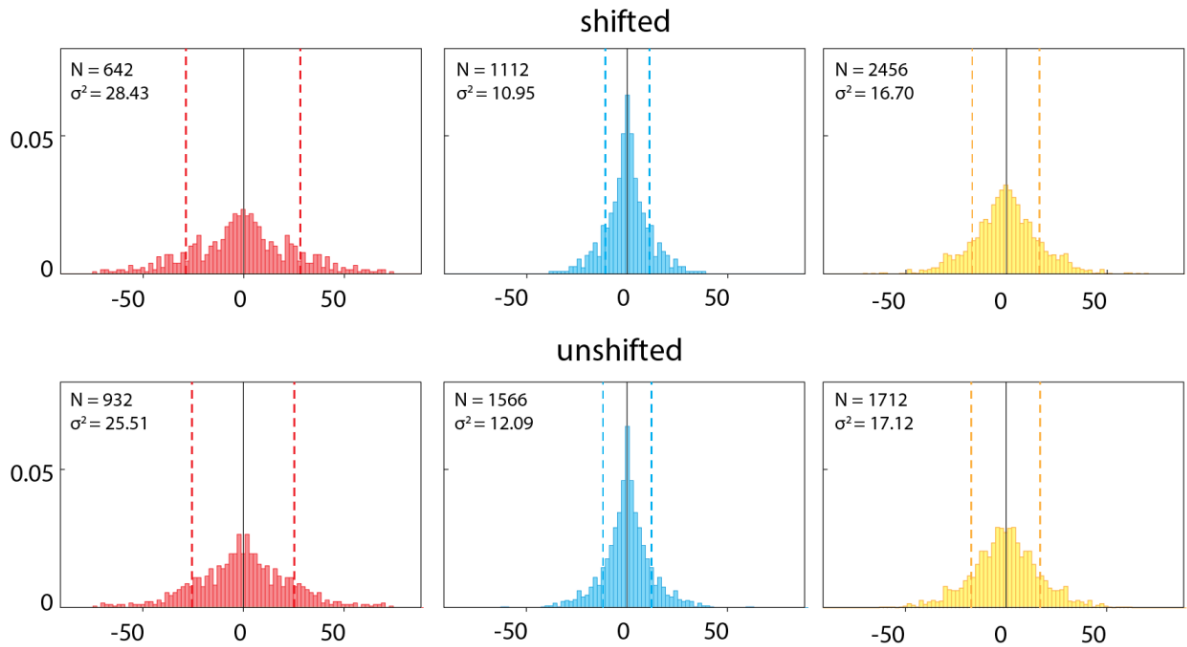

Figure 5.3. Turning angle PDF's (symmetrized) for the high-variance (red), low-variance (blue), and dominant (yellow) morphotypes. N is the number of turning angles in the symmetrized dataset (collected values = N/2), $\sigma^2$ is the variance.

Table 5.2. Percent of bioturbated grid squares belonging to each morphotype group.

|        | YELLOW | BLUE | RED | UNASSIGNED |
|--------|--------|------|-----|------------|
| **SLAB A** | 33% | 10% | 28% | 28% |
| **SLAB B** | 38% | 29% | 5% | 28% |
| **SLAB C** | 12% | 12% | 12% | 65% |
| **SLAB D** | 51% | 0% | 17% | 32% |

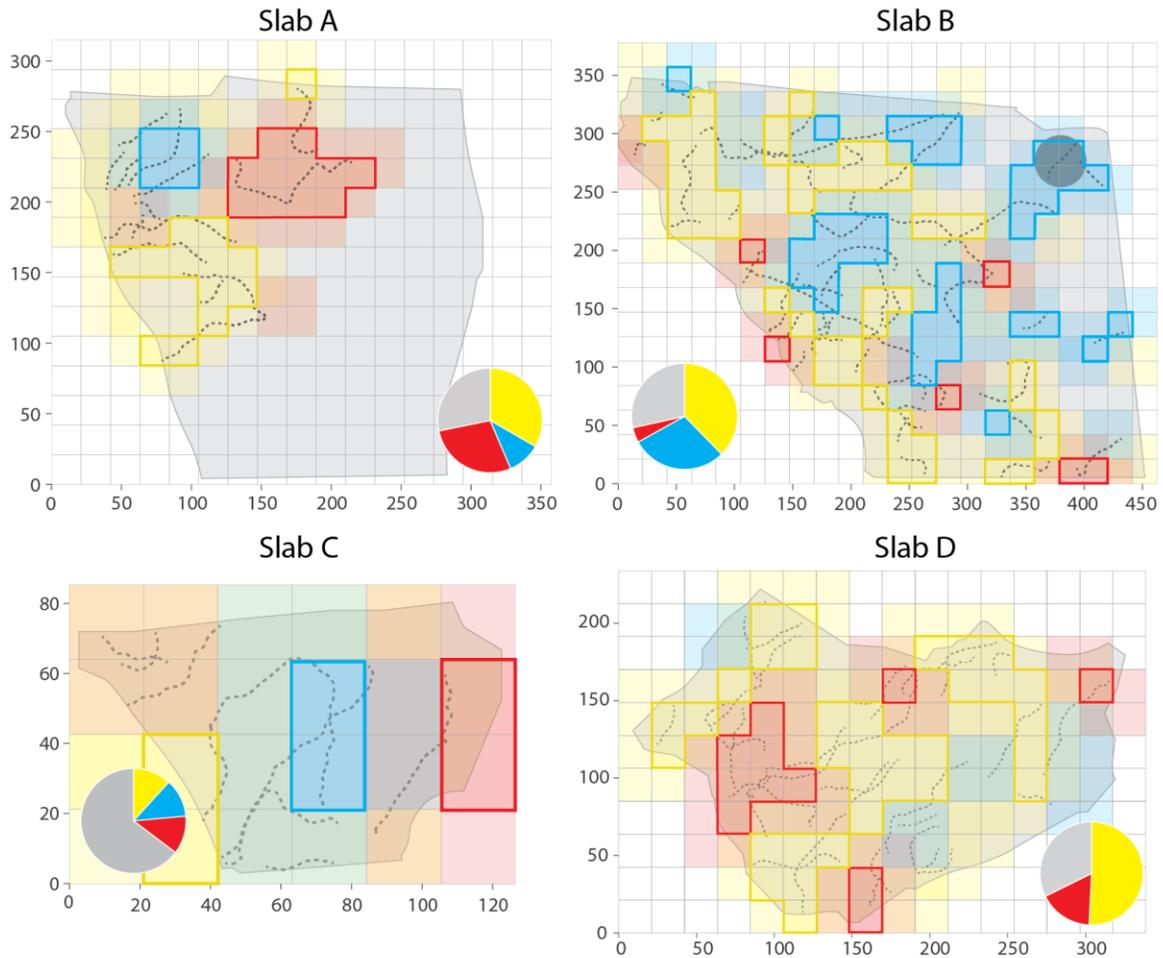

Figure 5.4. Results of spatial analysis of slabs containing *Helminthoidichnites tenuis* from the Ediacara Member, South Australia. Axes are in mm, slab extent denoted by grey shading, with trails indicated by dashed lines. Candidate grid squares denoted by either yellow (dominant morphotype), blue (low-variance subordinate morphotype), or red (high-variance subordinate morphotype). Candidate regions are denoted with a thick outline. Inset pie charts reflect the percent of bioturbated grid squares belonging to morphotype region, with grey indicating unassigned but bioturbated grid squares. Each grid square denotes an area of 21 mm by 21 mm (15 w by 15 w). Dark grey circle is the outline of a Dickinsonid fossil.

## 5.6 DISCUSSION

Our results demonstrate the presence of spatially variable *Helminthoidichnites tenuis* morphotypes on the studied Ediacara Member slabs. These regions are characterized by the presence of one of three behavioural variants: a dominant stereotyped behaviour, a more linear subordinate behaviour, and a more tortuous subordinate behaviour (i.e. yellow, blue, and red in Figure 5.4 respectively). Specimens demonstrate morphological variability along their trajectories, indicating the *Helminthoidichnites tenuis* tracemaker varied its behavioural strategy over time and space (i.e. behavioural flexibility *sensu* Wainwright et al., 2008). The presence of spatially distinct regions which contain more than one specimen suggests the variation in behavioural strategy may be explained, at least in part, by spatially restricted external factors. This indicates (1) that the *Helminthoidichnites tenuis* tracemaker was able to detect and respond to external cues, and (2) that these cues were spatially heterogeneous in the studied slabs. This is consistent with other statistical examinations which have demonstrated spatial aggregation in sessile and mobile macrofauna in nearshore Ediacaran environments, including *Funisia* from the Ediacara Member and trace fossils from the Shibantan Member of South China (Mitchell et al., 2020; 2022).

In the absence of preserved sensory organs in the Ediacaran, it is likely that the *Helminthoidichnites tenuis* tracemaker employed chemoreception to assist in navigation. In multicellular organisms, this sensory modality enables the organism to detect gradients of chemicals relevant to their survival such as total organic carbon, oxygen, or sulfur (Kirkegaard et al., 2016). In turn, almost all motile organisms exhibit movement in response to a stimulus. These responses can be *stochastic*, wherein the organism increases or decreases the frequency of undirected re-orientations in response to stimulus gradients, or *deterministic*, wherein the organism directs its re-orientation (i.e. steers) towards or away from stimulus gradients (Kirkegaard et al., 2016; Wan & Jékely, 2021). In either case increased number of re-orientations are an indication that the stimulus gradients are becoming increasingly unfavourable.

As a result, more linear paths form as an organism follows, or turns less often as it follows, favourable stimulus gradients, and more tortuous paths form as the organism turned, or turned more often, in search of favourable stimulus gradients or to avoid unfavourable ones (Pierce-Shimomura et al., 1999; Iino & Yoshida, 2009; Kirkegaard et al., 2016). If our results are analyzed in this light, then different morphotype regions may indicate areas where the organism altered its stereotyped behaviour (yellow candidate regions) in response to favourable (low variance = blue candidate regions) or unfavourable (high variance = red candidate regions) external stimulus gradients.

Even with simple chemoreception organisms detect and respond to various chemicals (e.g. oxygen, sulfur, total organic carbon, acidity) (Kirkegaard et al., 2016; Wan & Jékely, 2021). Thus, the behavioural heterogeneity observed in the studied slabs will be a combined result of the organism's internal state and cost-benefit analysis of negative and positive stimuli. Nutrient-rich regions may be unfavourable due to low oxygen concentrations or high acidity; or increase in favourability in times of extreme hunger. Likewise, nutrient-poor regions may be favourable due to high oxygen concentrations. As a consequence, it is unlikely that our results can be wholly explained by any one potential stimuli. Dissolved oxygen concentration in Ediacaran oceans, while greater than preceding times, are thought to have been spatially variable and markedly lower than modern levels (Canfield et al., 2007; Partin et al., 2013; Reinhard et al. 2013; Sperling et al., 2015; Evans et al., 2018; Wei et al., 2023). With higher oxygen and nutrient demands than sessile fauna, oxygen and nutrient availability may have served as two primary controls on the behaviour, and resulting movement trajectory, of the *H. tenuis* tracemaker (Mitchell et al., 2022.

Microbial mats may serve as important oxygen and nutrient hotspots in Ediacaran environments. *Helminthoidichnites tenuis* is well-accepted as the trace of a mat-grazing organism (Hofmann and Mountjoy, 2010; Carbone & Narbonne, 2014; Buatois et al., 2014; Gehling & Droser, 2018). Travel within microbial mats may additionally benefit the tracemaker due to co-occurring high oxygen concentrations caused by microbial respiration (Gingras et al., 2011; Scott et al., 2019). As a result, the distribution of such mats would undoubtedly impact the tracemaker's behaviour. The heterogeneous behaviour shown by our results could therefore be simply explained by heterogeneous distribution of microbial mats, causing variations in both oxygen and nutrient availability.

Ediacaran macrofauna could also serve as nutrient hotspots, if the *H. tenuis* tracemaker was able to exploit this food source (Budd & Jensen, 2017; Gehling & Droser, 2018). Only one dickinsonid body fossil occurs on the studied slabs (Figure 5.4, Slab B), providing limited insight for or against a scavenging hypothesis for the *H. tenuis* tracemaker. However, the method presented here may be utilized to further test such associations and claims that Ediacaran carcasses provided rich nutrient sources for mobile fauna.

Oxygen availability may be negatively impacted by decay-mediated hypoxia, produced by buried microbial mats or decaying macrofauna (Budd & Jensen, 2017). Thin, linear layers of sand formed under the influence of wave action are well documented from the Ediacara Member (Tarhan et al., 2017). Sand deposition could have smothered microbial mats, creating linear zones of limited oxygen availability. In fact, restriction of *H. tenuis* to regions with less than 15 mm of sedimentary overburden has been previously reported in the Ediacara Member (Gehling & Droser, 2018; Evans et al., 2020). In the slabs studied here, a possible trend in the occurrence of low-variance and dominant morphotype regions occur in Slabs B and D, trending to the top-left and top-right respectively and may reflect similar unfavourable zones underlying linear sand bodies (Figure 5.4).

The presence of spatially variable behaviour in *H. tenuis* in the Ediacaran Member indicate the tracemaker was able to detect and respond to spatially heterogeneous external cues. This could be via a stochastic chemotactic strategy, such as a biased random walk, or via a deterministic chemotactic strategy, such as contact chemoreception. However, there is a lack of distinct resource-focused feeding strategies (e.g. tight meanders, boundary following, net-like burrow networks) which would be expected with strongly heterogeneous nutrient landscapes and deterministic strategies, as seen in Cambrian ecosystems (Wang et al. 2009; Mángano et al., 2012, 2019; Zhao et al. 2018). These results agree with other spatial analyses on Ediacaran and early Cambrian paleocommunitiesm which suggest both sessile and mobile organisms responded to seafloor heterogeneity in nearshore environments; though Ediacaran ichnotaxa displayed limited resource-focused feeding when compared to Cambrian ichnotaxa (Mitchell et al., 2020; 2022).

Three possible scenarios for this paucity of resource-focused feeding strategies in the Ediacaran exist. (A) The environment was not heterogeneous enough: signal strength between favourable and unfavourable regions was not strong enough to trigger a deterministic strategy, but the capabilities for such strategies existed. (B) Perceptual range was limited, resulting in inefficient navigational capacities; or there was an absence of a nerve-sensory system able to support deterministic behaviours. (C) Combination of limited environmental heterogeneity and navigational capabilities: the environment was not strongly heterogeneous, nor were sensory and navigational capabilities sensitive enough to produce deterministic strategies. We suggest the latter "middle ground" scenario, involving a combination of intrinsic and extrinsic factors. Matgrounds may have displayed incipient heterogeneity; enough to produce some spatially variable behaviour even with limited navigational capacity (**Chapter 4**). This phenomenon may have triggered an early step in the positive feedback loop between environmental heterogeneity and the evolution of nerve systems with increased sensory and navigational capacities (i.e. the Cambrian Information Revolution) (Plotnick, 2010).

## 5.7 CONCLUSION

Here we present a novel methodology to investigate spatial trends in horizontal locomotory or grazing trace fossils. We compared fossil movement path morphology across regions and revealed three distinct and spatially restricted morphotypes in *Helminthoidichites tenuis* from the Ediacara Member interpreted to reflect a dominant stereotyped behaviour, a subordinate behaviour in response to favourable external conditions, and a second subordinate behaviour in response to unfavourable conditions. This demonstrates the *Helminthoidichnites tenuis* tracemaker was able to vary its behavioural strategy to external conditions and that these conditions may have been heterogenous on a centimeter scale. The potential causes of favourable or unfavourable conditions are numerous and difficult to determine with certainty, but could include heterogeneity in nutrient concentration, oxygen availability, or sediment overburden. Our analysis reveals an as-yet undiscovered variability in the behaviour of the *Helminthoidichnites tenuis* tracemaker and confirmed heterogeneity in external conditions relevant to the tracemaker in the latest Ediacaran.

## 5.8 ACKNOWLEDGMENTS

After the cancellation of B.A.L.'s field seasons due to the Covid-19 pandemic, Prof. Mary Droser graciously supplied field photographs of *Helminthoidichnites tenuis*, without which this paper would not be possible. B.A.L. was supported by a NSERC PGS-D grant and Macquarie IMQRES scholarship. Research by M.G.M. and L.A.B. was supported by Natural Sciences and Engineering Research Council (NSERC) Discovery Grants 311727–20 and 422931-20, respectively. M.G.M. thanks additional funding by the George J. McLeod Enhancement Chair in Geology.

## 5.10 SUPPLEMENTARY INFORMATION

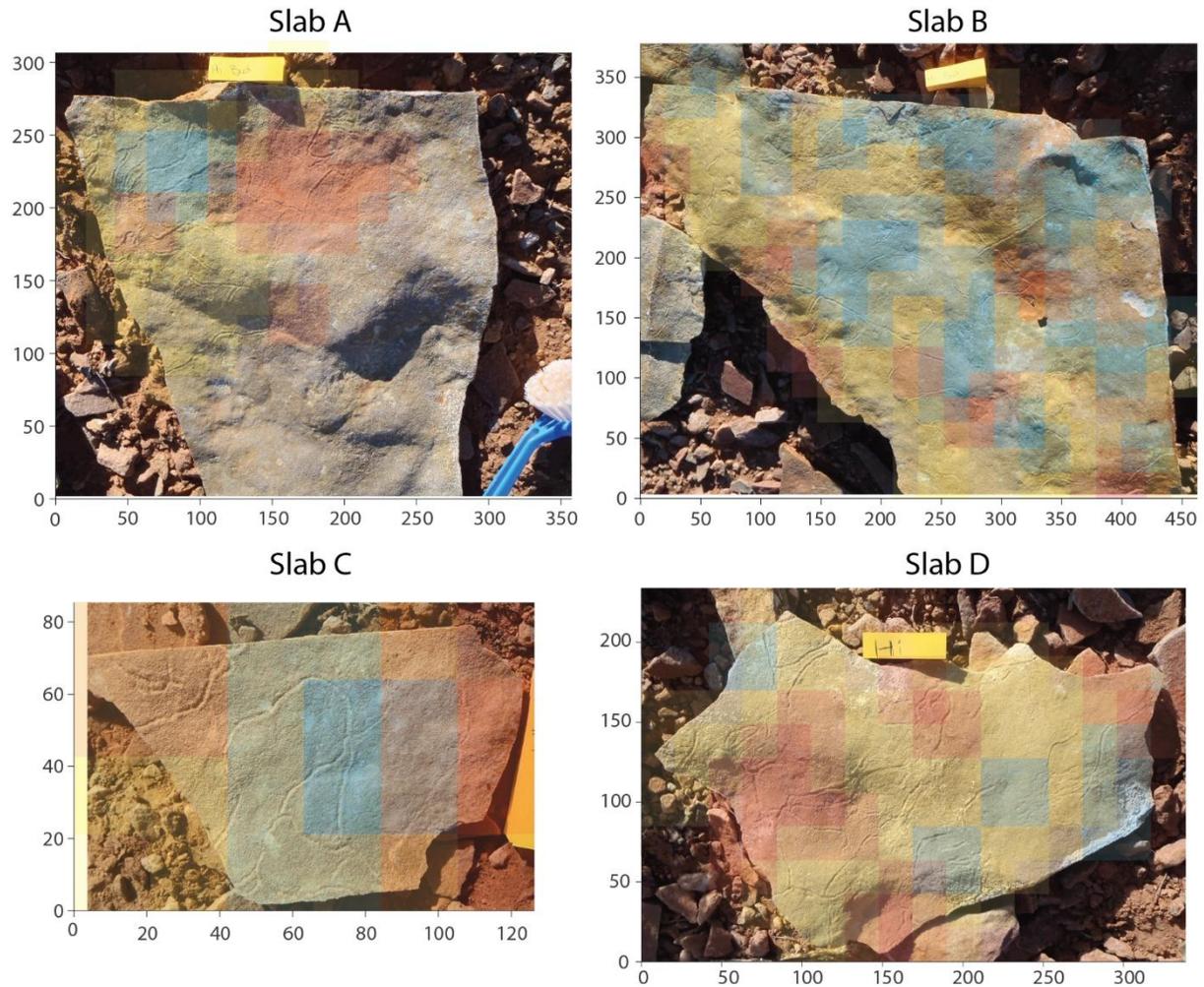

Figure 5.5. Results of spatial analysis of slabs containing *Helminthoidichnites tenuis* from the Ediacara Member in Southern Australia. Transparent colour denotes the morphotype present in each 42 mm by 42 mm (30 *w* by 30 *w*)analysis, which are then superimposed (yellow = dominant morphotype, blue = low-variance subordinate morphotype, or red = high-variance subordinate morphotype). Axes are in mm.

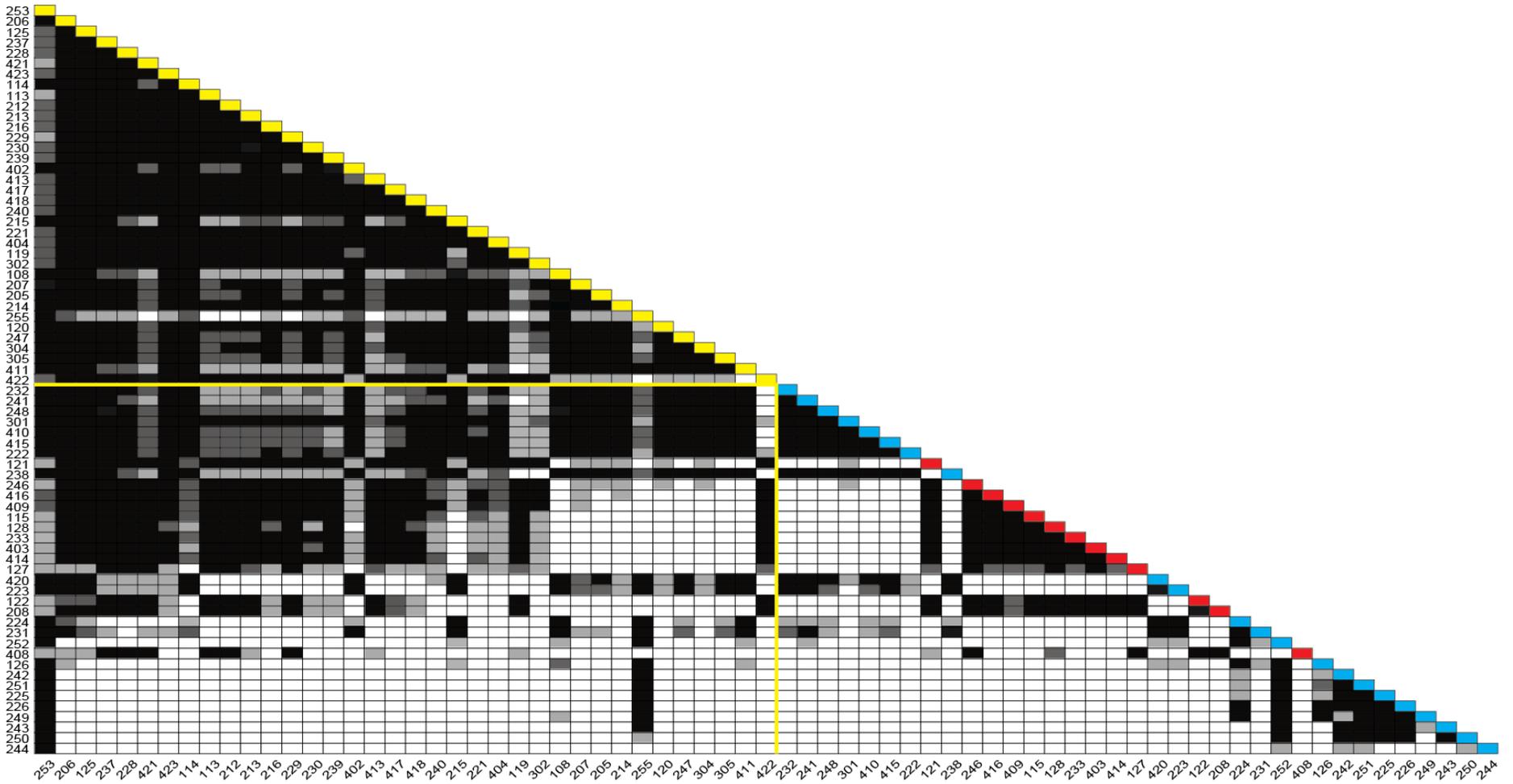

Figure 5.6. Results of two-sample f-tests comparing turning angle distributions from all populated unshifted grid squares. Grid square numbering shown in Figure 5.5. Yellow lines indicate extent of thresholding argument. Coloured squares indicate assigned morphotype.

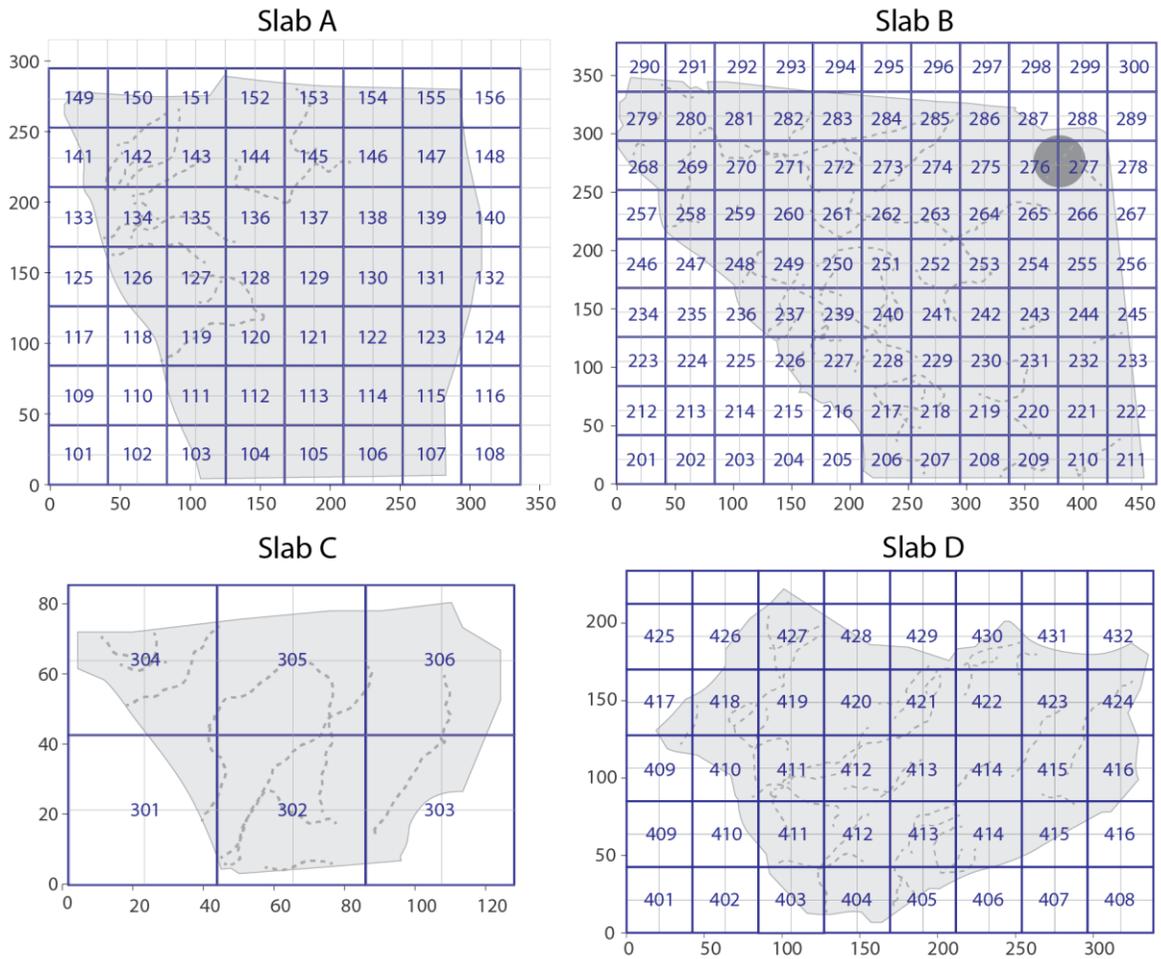

Figure 5.7. Grid square numbering used in two-sample f-tests, results shown in Figure 5.6.